\documentstyle[preprint,epsf]{jpsj}
\title{\bf Griffiths-McCoy singularities in the transverse field 
Ising model on the randomly diluted square lattice}
\begin{document}
\date{\it{\today}}
\author{Tohru Ikegami and Seiji Miyashita\\
{\it Department of Earth and Space Science, 
                        Graduate School of Science, Osaka University}\\
\smallskip
Heiko Rieger\\
{\it HLRZ, Forschungszentrum J\"ulich, 52425 J\"ulich, Germany}
}
\abst{
The site-diluted transverse field Ising model in two dimensions is
studied with Quantum-Monte-Carlo simulations. Its phase diagram is 
determined in the transverse field ($\Gamma$) and temperature ($T$) plane 
for various (fixed) concentrations ($p$). 
The nature of the quantum Griffiths phase at zero temperature is 
investigated by calculating the distribution of the local 
zero-frequency susceptibility. 
It is pointed out that the nature of the Griffiths phase is different 
for small and large $\Gamma$.
}
\kword{\rm \ QPT;\ Griffiths-McCoy singularities;\ multi-critical point;\ 
dynamical exponent 
}
\sloppy
\maketitle
\newcommand{\bc}{\begin{center}}
\newcommand{\ec}{\end{center}}
\newcommand{\be}{\begin{equation}}
\newcommand{\ee}{\end{equation}}
\newcommand{\beqn}{\begin{eqnarray}}
\newcommand{\eeqn}{\end{eqnarray}}

\section{Introduction}

In numerous investigations of the effect of quantum fluctuations on the 
order-disorder phase transition the transverse-field spin-$1/2$ Ising model 
with ferromagnetic interactions has served as the simplest model on which 
these effects can be studied \cite{bikas}. This model is of particular 
interest, since the strength of the quantum fluctuations can be 
controlled by varying the strength $\Gamma$ of the transverse field: a pure 
quantum phase transition (QPT) occurs at zero temperature when the external 
field exceeds a critical value. 
The necessary prerequisite for the existence of long range order 
is the existence of an infinite connected component or cluster in the
lattice, because only an infinite system can have a spontaneous
magnetization. 
As has been recognized by Harris in the context of randomly site diluted 
lattices an infinite connected cluster is not only necessary but also 
sufficient for the existence of such a QPT. Any percolating cluster is at 
least one-dimensional (i.e.\ the fractal dimension of its backbone is larger 
than one). Since even the one-dimensional transverse field Ising model shows a 
QPT at some non-vanishing strength of the transverse field, any infinite 
cluster must have a QPT.

The investigation of the effects of randomness on the transverse field
Ising models (TFI) has attracted a lot of interest recently 
\cite{review}. In particular, the peculiar features of random systems which 
derive from the Griffiths-McCoy singularity have been studied intensively 
\cite{griffiths,mccoy}. 

As Griffiths has pointed out \cite{griffiths}, randomly diluted
(classical) ferromagnets have a non-analytic free energy even away
from the thermodynamical critical temperature for a given
concentration $p$.  As a matter of fact for any fixed concentration
$p$ of occupied sites such a non-analytic behavior persists up to a
temperature which coincides with the transition temperature of the
pure case ($T_c$ for $p=1$). The reason for these anomalies is that
there is a non-vanishing probability to find arbitrarily large
clusters of connected spins for any concentration $p$.  Since large
clusters tend to order ferromagnetically below $T_c$, they act very
coherently. This causes singularities in the response to an external
field and in the dynamical properties even above the critical point.
These effects have been pointed out as significant effects in the
distribution of zeros of the partition function and the Laplace
transform of dynamical correlation \cite{green}.

The Griffiths phase persists in the presence of a
transverse field and is, for fixed concentration $p$, located between
the ferromagnetic phase boundary $T_c(\Gamma;p)$ and the pure paramagnetic 
phase boundary 
$T_c(\Gamma;p=1)$. While the effect of the Griffiths--McCoy singularity on 
the relaxation process of the diluted Ising model causes a non-exponential 
decay of the time correlation function \cite{RS,Bray,takano}, it has no 
effect on the thermodynamic properties for non-zero temperatures. At zero 
temperature, however, the singularity causes singular behavior even in static 
properties \cite{fisher,thill,rieger2,guo2,rieger3}. Particularly noteworthy 
are the divergence of the local susceptibility and the algebraic decay of the 
dynamical correlations. 

In this paper we study the transverse field spin-1/2 Ising model on a
randomly diluted square lattice. Here, we mainly investigate the
distribution of the local susceptibility using the continuous
(imaginary) time cluster algorithm for transverse Ising models
developed by Kawashima and Rieger \cite{con1} (continuous time
cluster algorithms for various other quantum mechanical models have
been proposed recently \cite{con2}). It turns out that this
distribution has a power law decay for large values, and consequently
the local susceptibility diverges at some points in the disordered
phase. A similar behavior has been found for quantum spin glass
systems, but for these systems only the local nonlinear
susceptibilities diverge \cite{rieger2,guo2}.  The exponent for the
power law decay of the distribution is related to the dynamical
exponent $z$. As our most important result we confirm the result of
Senthil and Sachdev \cite{senthil} that at the quantum critical point
time scales diverge exponentially fast at the percolation threshold,
implying that the dynamical critical exponent $z_{crit}$ is infinite, 
whereas in 2 or 3-dimensional transverse field Ising spin glasses a
finite value for $z_{crit}$ has been estimated \cite{rig}.  The
concentration dependence of the phase boundary in the site or bond
diluted square lattices at zero temperature is also of interest.
Particularly interesting features of the phase diagram such as the
existence of a multi-critical point and a straight vertical phase
boundary have been observed for small transverse fields
\cite{harris,stinchcombe,ray}. For such small transverse fields the
percolation threshold determines the boundary between the ordered
(ferromagnetic) phase and the disordered phase.  On the other hand for
large transverse field, even in the geometrically connected cluster
(i.e. above the percolation threshold), the magnetically connected
cluster is reduced by quantum fluctuations.

\section{Model and Method}

The model that we consider here is the spin-1/2 Ising model in a
transverse field on a square lattice with random site dilution, i.e.\
sites on a square lattice are occupied with spins with probability $p$
and empty with probability $1-p$. Only occupied nearest neighbor
sites interact. The model is thus defined by the quantum mechanical
Hamiltonian
\be
H = -J\sum_{\langle i,j\rangle}
      \varepsilon_i\varepsilon_j\sigma_i^z\sigma_j^z
    -\Gamma\sum_i\varepsilon_i\sigma_i^x\;,
\label{hamil}
\ee
where $\sigma_i$ are Pauli spin matrices, $\langle i,j\rangle$ denotes
nearest neighbor pairs on a $L\times L$ square lattice with periodic
boundary conditions, $J$ is the ferromagnetic interaction strength and
$\Gamma$ is the strength of the transverse field; $\varepsilon_i\in\{0,1\}$ 
are quenched random variables modeling the dilution: $\varepsilon_i=1$ with 
probability $p$ and $\varepsilon_i=0$ with probability $1-p$. 

To map the 2-dimensional quantum system to the 3-dimensional classical 
system, we use the Su\-zu\-ki-Trotter decomposition \cite{suzuki}. Then 
the free energy ${\cal F}$ of the system (\ref{hamil}) at inverse temperature 
$\beta=1/T$ is obtained as the limit of a 3-dimensional classical Ising 
model:
\be
{\cal F} =  -\beta^{-1}\lim_{n\to\infty}
\ln\,{\rm Tr}\exp(-{\cal S_{\rm class}})\;,
\label{free}
\ee
with
\be
{\cal S_{\rm class}} = 
-K_{hor}\sum_{\tau=1}^n\sum_{\langle ij\rangle} \varepsilon_i\varepsilon_j
S_{i}(\tau) S_{j}(\tau)
-K_{ver}\sum_{\tau, i} \varepsilon_i S_{i}(\tau) S_{i}(\tau+1)\;,
\label{class}
\ee
and
\be
\begin{array}{lcl}
K_{hor}     & = & \Delta \tau J \;,\\
K_{ver}     & = & -\frac{1}{2}\ln\tanh(\Delta \tau \Gamma) \;,\\
\Delta \tau & = & \beta/n \;.
\end{array}
\ee
The classical action (\ref{class}) is the Hamiltonian of a cubic
lattice of $L\times L\times n$ classical Ising spins
$S_i(\tau)=\pm1$. The additional index $\tau=1,\ldots,n$ labels the
$n$ two-dimensional (imaginary) time slices within which spins
interact via $K_{hor}$ and among which they interact with strength
$K_{ver}$.  The number of time slices $n$ is called the Trotter number
and the third (imaginary time) axis is called the Trotter axis.

Note that an empty (occupied) site $i$ in one two-dimensional time
slice implies a whole column of empty (occupied) sites in the Trotter
direction: the quenched disorder is perfectly {\it correlated} in the
imaginary time direction. Therefore the model we study is {\it not} a
diluted ferromagnet on a cubic lattice with {\it uncorrelated}
disorder, and its universality class will be different from that of the site 
diluted cubic ferromagnet. We obtain the phase diagram for the diluted model 
by the conventional quantum Monte Carlo method using the world line update 
for finite Trotter numbers, Fig. 1. The phase boundaries for 
$p=0.8$, $0.7$ and $0.6$ are also shown. For finite temperatures, 
the phase boundary shrinks as $p$ decreases towards $p_c$($ \simeq 0.59$). 
It is shown that the Griffiths phase for fixed concentration $p$ is located 
between the FM phase boundary $T_c(\Gamma;p)$ and the pure PM phase boundary 
$T_c(\Gamma;p=1)$. But the conventional method has difficulties in the regions 
close to $T=0$, because we need $n$ proportional 
to $\beta$ to take the Trotter limit explicitly. Moreover, it is extremely 
hard to equilibrate the system for weak transverse fields. In this work, 
we use the new continuous time cluster algorithm developed by Kawashima 
and Rieger \cite{con1} to avoid these difficulties. For a description 
of the algorithm and details of the implementation we refer to \cite{con1}.

First, we have checked our method for the pure case (i.e.\ $p=1$), for
which the phase diagram is already known and the critical field value
and the thermal exponents have been estimated from series expansions
\cite{el1,el2}.  As a test of the cluster algorithm, we plot 
the square of the magnetization $\left<M^2\right>$ obtained at each Monte 
Carlo step, Fig. 2. At high temperature and low $\Gamma$
the system is in the disordered region.  However, due to the strong
coupling constant along the Trotter axis, the time to reach
equilibration is large. The use of the cluster algorithm reduces this time 
significantly. In Fig. 3, we study $n$ dependence of the critical values of 
$\Gamma$ obtained by the Binder plot ( the dimensionless ratio of moments of 
the order parameter ) at $\beta=10$. The value obtained from the continuous 
cluster algorithm is approximately $3.05$. This is indeed close to the value 
which is obtained by extrapolating from the data of the conventional method.
Finally, we follow Kawashima and Rieger \cite{con1} and estimate the
critical $\Gamma$ at zero temperature and to obtain the critical
exponents from finite size scaling.

Close to the quantum critical point, quantities are expected to obey
the finite size scaling form
\be
\left<A\right>=L^a\tilde{q}\left(L^{1/\nu}\delta,L^z/\beta\right)\;,
\ee
where $\delta=\Gamma-\Gamma_c$ and $a$ is the finite size scaling
exponent of the quantity $A$. From the equivalence with the
3-dimensional classical Ising model one knows the dynamical exponent
to be $z=1$. Thus, we can perform conventional one-parameter finite
size scaling, if we choose the aspect ratio $\beta/L$ to be constant.
We work with a constant aspect ratio $\beta/L=1/5$ and estimate the
magnetization $m$ ($a=-\beta/\nu$), the uniform susceptibility $\chi$
($a=2-\eta$) and the Binder ratio $g$ ($a=0$) in order to determine
the values of $\nu$, $\eta$ and $\beta$, see Fig. 4. As a result, we
estimate $\Gamma_c\simeq 3.06$, $\nu\simeq 0.64$, and $\beta\simeq
0.31$ from $m$, $\Gamma_c\simeq 3.07$, $\nu\simeq 0.63$ and
$\eta\simeq 0.04$ from $\chi$, and $\Gamma_c\simeq 3.08$ and
$\nu\simeq 0.63$ from $g$. These values agree well with the series
expansion results \cite{el2} and with those obtained earlier by
Kawashima and Rieger \cite{con1} using the same method.

\section{Quantum Griffiths phase}

The phase boundary in the site diluted square lattices is depicted in 
Fig. 5. As has been mentioned above we expect the Griffiths phase in the 
classical model ($\Gamma=0$) between the critical temperature of the pure 
model $p=1$ and the thermodynamical critical temperature for the 
concentration $T_c(p)$. The singularity in this Griffiths phase is 
due to the finite probability of large clusters which behave 
coherently below $T_c$. For low concentration the probability 
is given approximately by
\be
P\left(N\right)dN \propto p^NdN = \exp\left(-N\left|\ln p \right|\right)\;,
\label{clusterprob}
\ee
i.e.\ it vanishes exponentially with the number of spins. On the other hand, 
at $T=0$ (quantum region) the existence of a vertical line (parallel to the 
$\Gamma$-axis) at $p_c$ (the percolation threshold $\sim 0.59$ \cite{ziff}) 
extending from $\Gamma=0$ to $\Gamma_M$ follows from the nature of the 
backbone of the percolating cluster \cite{harris}, which has a fractal 
dimension between 1 and 2. Therefore the critical transverse field strength 
of the pure one-dimensional transverse Ising model is a lower bound for 
$\Gamma_M$, i.e. $\Gamma_M\ge J=1$.

Again, due to the presence of arbitrarily large connected clusters 
the entire area below $\Gamma_c$ is Griffiths phase (quantum Griffiths phase) 
even the ordered phase contains anomalies. The paramagnetic phase lies 
above $\Gamma_c$. Since we are at zero temperature close 
to the quantum phase transition points $\Gamma_c(p)$, we encounter 
as a new feature much 
stronger singularities than in the classical case, a fact that has first 
been noted by McCoy \cite{mccoy} and has been intensely investigated recently 
in various situations \cite{fisher,thill,rieger2,guo2,rieger3}. 
The transition across 
the vertical line including the multi-critical point at 
$(0,1-p_c,\Gamma_M)$ as well as the Griffiths--McCoy singularity for small 
transverse fields $\Gamma<\Gamma_M$ have also been discussed quite 
recently \cite{senthil}.

The main aims of our present investigation are to determine the dynamical 
exponent in the quantum Griffiths phase and to confirm the conjecture 
of the straight vertical phase boundary. For $p\ll p_c$ the probability 
distribution of the existence of a spherical 
dense cluster with $N$ spins is given by Eq.(\ref{clusterprob}). In the 
transformed 3D classical Ising model, such a cluster forms a rod along the 
Trotter axis as shown in Fig. 6. The spins in the cluster are strongly coupled 
in real space and may have a domain wall perpendicular to the Trotter axis.
The upward or downward solid arrow indicates parallel or anti-parallel with 
respect to the state of the cluster in real space. The insertion of a domain 
wall costs an energy $\Delta E$
\be
\Delta E \simeq C N \;,
\label{stiffness}
\ee
where $C$ is stiffness constant. 
The probability for such a domain wall decreases exponentially as
$\exp(-CN)$. Therefore the correlation length along the Trotter axis as
a function of a cluster size $N$ is given by
\be
\xi_{\tau}\left(N\right) \sim \exp\left(CN\right).
\label{corrlength}
\ee
Now we introduce the local susceptibility $\chi_{local}$, which is the 
response of the expectation value of a spin $\sigma_i^z$ to a local 
(longitudinal) field $h_i$ on site $i$ which is expressed by  
$h_i\sigma_i^z$ in the Hamiltonian.
\be
\left.\frac{\partial}{\partial h_i}
\left\langle\sigma_i^z\left(h_i\right)\right\rangle\right|_{h_i=0}
=\chi_{local}\;.
\ee
In the continuous time method, following Kawashima and Rieger
\cite{con1}, the expectation value for the local susceptibility is
\be
\chi_{local}=\int_0^{\beta}\langle\sigma_i^z(\tau)\sigma_i^z(0)\rangle
=\beta\langle m_i^2\rangle\;.
\ee
Here, the local magnetization $m_i$ at site $i$ is the difference 
between the total length of all $+$segments and the total length of all 
$-$segments divided by $\beta$. From the relation (\ref{corrlength}), 
$\chi_{local}$ is proportional to $\xi_{\tau}$. Now let us consider the 
distribution of $\xi_{\tau}$, $P\left(\chi_{local}\right)$. The 
dependence of the 
correlation length $\xi_{\tau}$ on the cluster size $N$ is given by 
(\ref{corrlength}) and $N$ has the distribution (\ref{clusterprob}). 
Thus the distribution of $\xi_{\tau}$ is
\be
P\left(\xi_{\tau}\right)d\xi_{\tau}
\simeq\xi_{\tau}^{\vartheta}d\xi_{\tau}\;,
\ee
where
\be
\vartheta\equiv\frac{\ln p}{C}-1\;.
\ee
Consequently $P\left(\chi_{local}\right)$ has a power law dependence on 
$\chi_{local}$ at large values, 
\be
P\left(\chi_{local}\right)d\chi_{local} 
\simeq \left(\chi_{local}\right)^{\vartheta}d\chi_{local}\;.
\ee
The numerical estimation of the integrated distribution function
has a better statistics than $P\left(\chi_{local}\right)$. Therefore, in the 
practical calculation we obtain the integrated distribution function
\be
F\left(\chi_{local}\right)=
\int_{\chi_{local}}^{\infty}P\left(t\right)dt 
\sim \frac{C}{\ln p}\left(\chi_{local}\right)^{\ln p/C}\;.
\label{F}
\ee
In Fig. 7-11, the integrated distribution functions are plotted for 
various values of $\Gamma$, and $p$. We expect that the integrated 
distribution of $\chi_{local}$ will be cut off at $\chi_{local}=\beta$, and 
that increasing $\beta$ will simply extend the range over which the data 
fall on a straight line. 

Near the percolation threshold $p_c$, where the form (\ref{clusterprob}) 
is no longer valid, $P\left(N\right)$ is given by \cite{perc}
\be
P\left(N\right) \sim N^{-\tau}\exp\left(-\frac{aN}{\xi^D}\right)\;.
\label{clusterperc}
\ee
Here, $\xi$ is correlation length in real space which diverges as
$\xi \sim \left|p-p_c\right|^{-\nu_p}$, and D is the fractal dimension of 
the percolating cluster in two space dimensions.
Using Eq.(\ref{clusterperc}) instead of Eq.(\ref{clusterprob}), 
one obtains a formula for
$F\left(\chi_{local}\right)$ that is valid near the percolation
threshold as long as the phase boundary is vertical, i.e.\ 
$\Gamma\le\Gamma_{M}$:
\be
F\left(\chi_{local}\right) \sim 
\frac{C\xi^D}{a}\left(\chi_{local}\right)^{-a/C\xi^D}.
\label{lowgamma}
\ee
Following Rieger and Young \cite{rieger2} one can relate the power of the 
tail of $F(\chi_{local})$ to a single dynamical exponent 
$z(p,\Gamma)$ that varies continuously with $p$ and $\Gamma$,
\be
\ln F(\chi_{local}) = -\frac{d}{z}\ln\chi_{local} 
+ {\rm const.}\;,
\label{RY}
\ee
where $d$ is the space dimension, i.e.\ $d$=2 here. From (\ref{F}), 
(\ref{lowgamma}) and (\ref{RY}) it follows that 
\be
\frac{d}{z}=\left\{
\begin{array}{clcl}
-\ln p/C  & \quad{\rm for}\quad p\ll p_c\\
 a/C\xi^D & \quad{\rm for}\quad p\to p_c\:.
\end{array}
\right.
\ee
The second relation implies that $z$ diverges at $p_c$ algebraically 
\cite{senthil} like
\be
z\propto\vert p-p_c\vert^{-D\nu_p} \quad{\rm for}\quad 
p\to p_c\quad{\rm and}\quad\Gamma\le\Gamma_{M}.
\label{dz}
\ee
In Fig. 11 the data for $d/z$ are plotted for various values of $\Gamma$, and 
$p$. For $\Gamma=0.7,1.0$ ($\le\Gamma_M$), the data for $d/z$ vanish at 
$p_c$, whereas for $\Gamma=1.5,2.0$, they vanish for $p>p_c$. 
For $\Gamma>\Gamma_{M}$ the phase boundary is 
no longer parallel to the $\Gamma$-axis, and the quantum fluctuations are 
strong enough to destroy the ferromagnetic order along the backbone of a 
percolating cluster at $p_c$. 
Thus, our estimates for $d/z$ should be non-zero at 
$p_c$ for large transverse field strength $\Gamma>\Gamma_{M}$, since we 
expect the dynamical exponent $z$ to diverge only at 
$p_{\rm FM}(\Gamma)$, where $p_{\rm FM}(\Gamma)$ is the critical 
concentration at transverse field strength $\Gamma$. The observation of 
a nonzero slope at $p_c$ implies that $p_c$ is not a special
concentration for $\Gamma>\Gamma_{M}$.

Now let us investigate the criticality of $z$ (\ref{dz}) for 
$\Gamma<\Gamma_{M}$. The value of $D\nu_p$ obtained from the percolation 
theory is known as $D\nu_p\simeq 2.57$. As is shown in Fig. 12, the 
criticality of $d/z$ seems to be indeed compatible with the relation 
(\ref{dz}). For $\Gamma>\Gamma_{M}$ the phase boundary is no longer 
parallel to the $\Gamma$-axis and the quantum fluctuations are strong enough 
to destroy the ferromagnetic order on the backbone of a percolating cluster 
at $p_c$. We also analyzed the criticality of $z$ for $\Gamma>\Gamma_M$. 
We do not know, however, the critical value of $p_{FM}\left(\Gamma\right)$. 
Thus we tried to fit the data in the form 
$\vert p-p_{FM}\left(\Gamma\right)\vert^{-a}$. Minimizing the deviaton from 
the fitting form, we obtain $p_{FM}\left(1.5\right)\simeq 0.67$ and 
$a\simeq 2.36$, and $p_{FM}\left(2.0\right)\simeq 0.77$ and $a\simeq 2.24$. 
The values of $a$ are not far from for $\Gamma<\Gamma_M$. However, in order 
to conform the same universality class for the two regions, we need a more 
precise investigation which will be reported elsewhere. 

\section{Discussion and Summary}

The nature of the quantum Griffiths phase of the diluted transverse field 
Ising ferromagnet in two dimensions has been studied.  Due to the 
Griffiths-McCoy singularity, the local susceptibility $\chi_{local}$ 
diverges, if $d/z$ is smaller than 1. Thus, as in the case of a quantum spin 
glass \cite{rieger2,guo2}, the Griffiths-McCoy singularity 
causes a divergence of various static quantities such as the zero-frequency 
local susceptibilities. In the classical case the Griffiths singularity is 
not strong enough to produce such a dramatic effect.

So far we assume that the clusters are well defined geometrically and
we have used the probability function $P\left(N\right)$ given by percolation 
theory. However, even in the geometrically connected cluster the 
correlation function is reduced due to the quantum fluctuations, and 
when $\Gamma$ becomes large we have to work with the probability 
distribution of ``{\it physical clusters}'' which are smaller than the 
geometrical ones. Although 
we do not know $P\left(N_{phys}\right)$ as a function of $\Gamma$ at this 
moment, generally we expect to find the following senario:

For $\Gamma < \Gamma_M$, $P\left(N_{phys}\right) \simeq
P\left(N\right)$, and the mean cluster size diverges at $p=p_c$. On
the other hand for $\Gamma > \Gamma_M$, the mean size of the physical
cluster diverges at $p_{FM}\left(\Gamma\right) > p_c$. In Fig. 11, we
find that $d/z$ vanishes at $p_c$ for the values $\Gamma=$0.7 and 1.0,
which are both smaller than $\Gamma_{M}$. On the other hand, for
$\Gamma > \Gamma_M$ they vanish at a larger value of $p$, which is
considered to be $p_{\rm FM}(\Gamma)$. For $\Gamma>\Gamma_M$ it is 
difficult to estimate the exponent correctly. Because the phase boundary at 
the zero temperature is not yet known correctly. The question of whether the 
critical exponent of this physical cluster may be different from the value 
of the exponent in the regions $\Gamma<\Gamma_M$ is still open.

For $\Gamma \simeq \Gamma_c$ quantum fluctuation are 
very strong. In this case the present analysis which makes use of the broad 
distribution of $\chi_{local}$ does not apply for the small systems which 
we are able to investigate.  The correlation function along the Trotter
axis is not well developed, and the relation $\xi_{\tau} \sim 
\exp\left(CN\right)$ is no longer valid. 
The nature of this region which is dominated by quantum fluctuations will be 
studied elsewhere.\\
\\
{\bf Acknowledgements:} The authors would like to thank Professor Hiroshi 
Takano for his valuable comments and discussion, and to Professor H. -U. 
Everts for his critical reading, and also acknowledge fruitful assistances 
from M. Kikuchi and G. Chikenji in developing an efficient computer code. 
H.R.'s work was supported by the Deutsche Forschungsgemeinschaft (DFG) and 
he acknowledges gratefully financial support from the 
Japan-German (JSPS-DFG) cooperation project JAP-113, by which a fruitful 
stay, contributing essentially to the present work, at the Osaka university 
has been made possible. 
The present work is partially supported by Grant-in-Aid from the Ministry of 
Education, Science and Culture. They also appreciate for the facility of 
Supercomputer Center, Institute for Solid State Physics, University of Tokyo.

\newpage

\begin{figure}
\epsfxsize=16cm\epsfysize=15cm\epsfbox{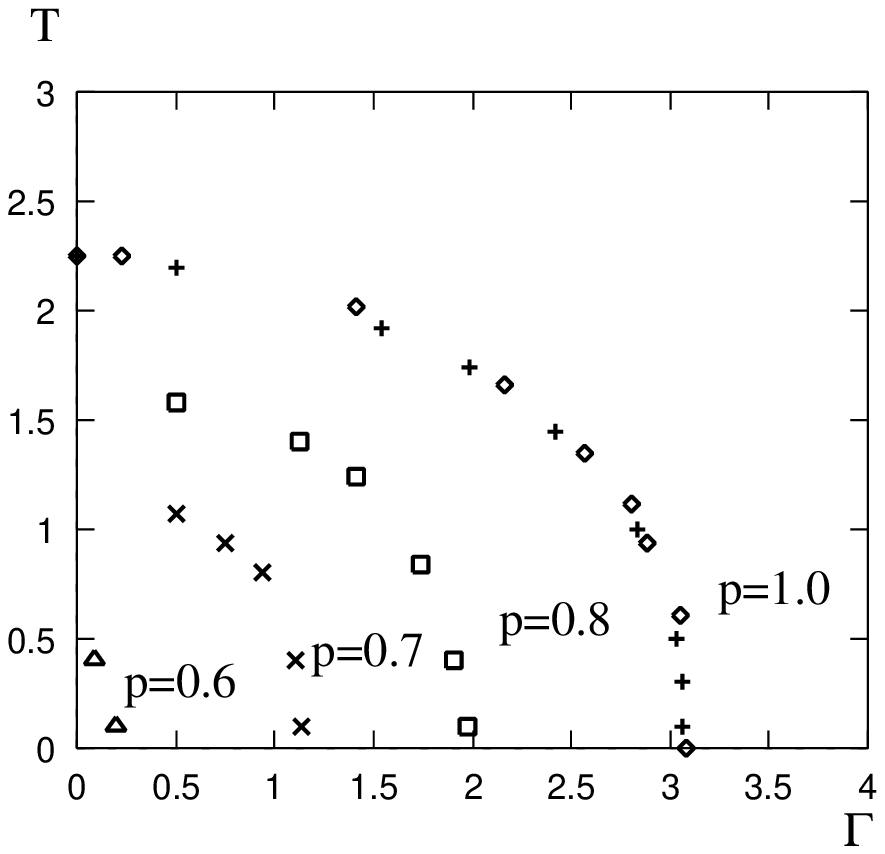}
\caption{
\label{fig1}
Phase diagram of the transverse field Ising model on the 
square lattice for the finite temperatures. The symbols 
$\left(\diamond\right)$ denote the results of the series expansions by 
Elliott and Wood, and the symbols $\left(+\right)$ denote the values obtained 
from the conventional method. In the diluted system. The critical line 
shrinks as $p \rightarrow p_c$.}
\end{figure}

\begin{figure}
\epsfxsize=16.5cm\epsfysize=14cm\epsfbox{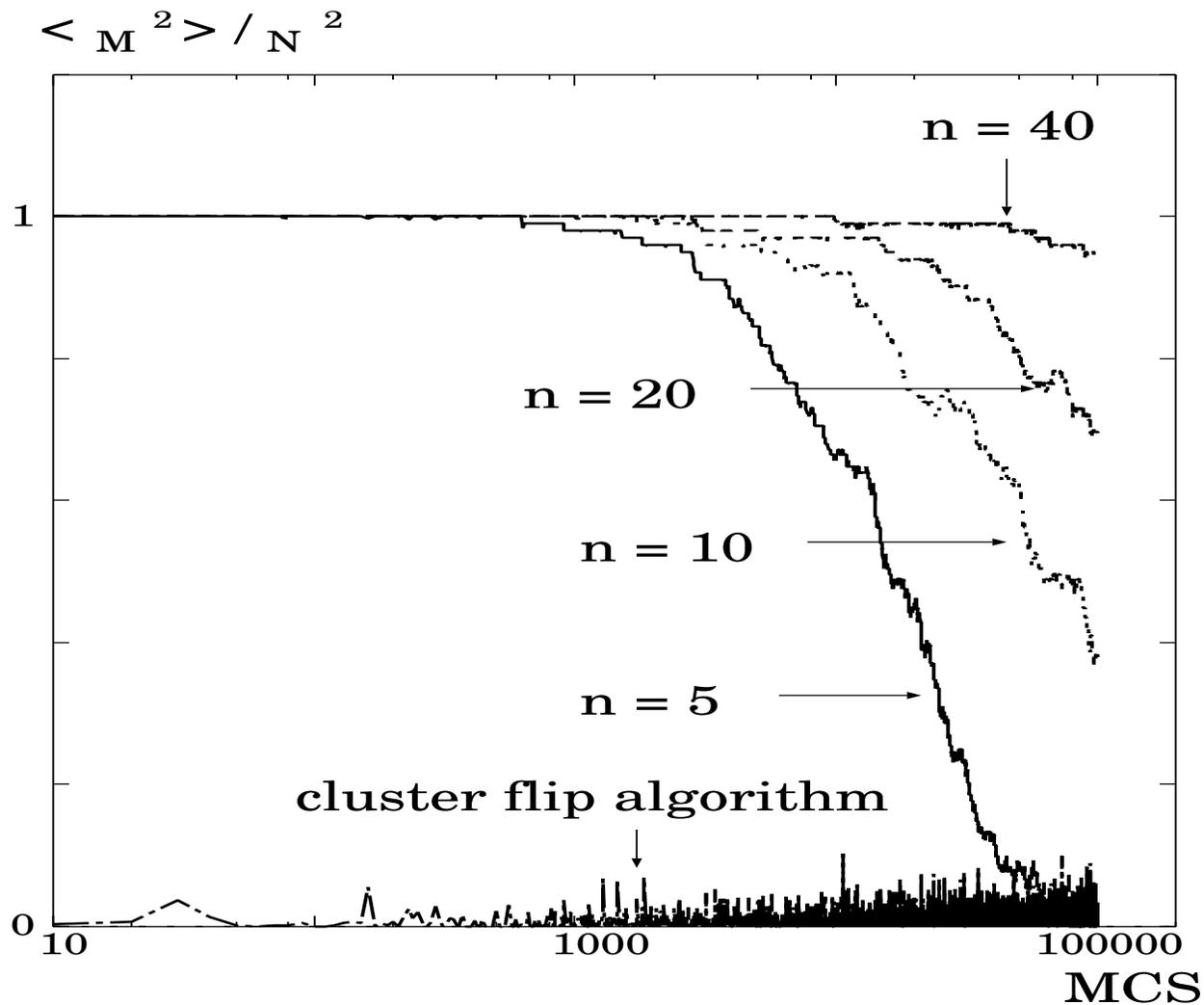}
\caption{
\label{fig2}
The Monte Carlo sweep dependence of the order parameter $\left<M^2\right>$ 
at $\beta=0.2$, $\Gamma=0.1$ starting from the complete ferromagnetic state.} 
\end{figure}

\begin{figure}
\epsfxsize=16cm\epsfysize=15cm\epsfbox{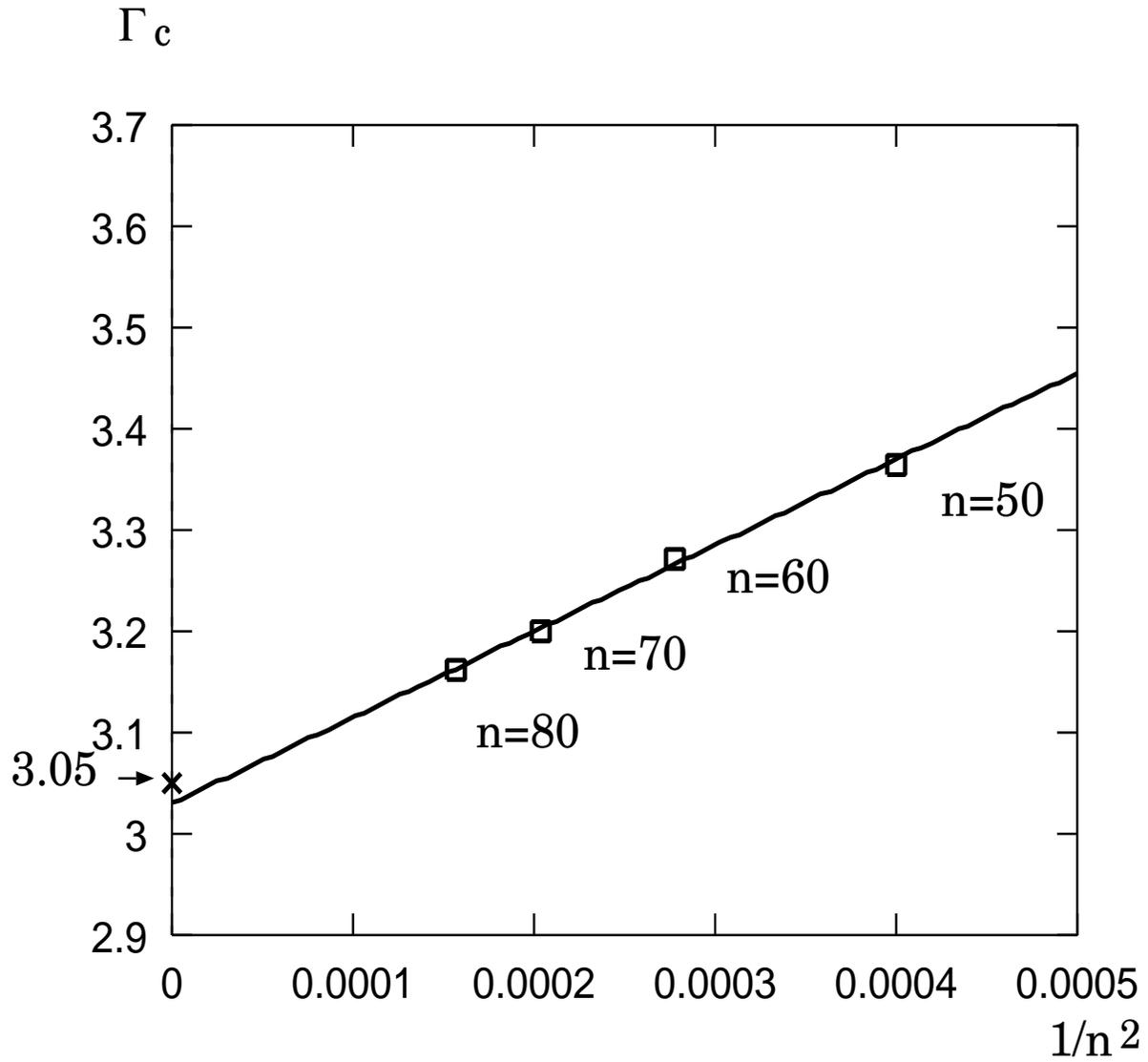}
\caption{
\label{fig3}
Estimation of the critical $\Gamma$ at $\beta=10$. The cross 
($\Gamma_C\simeq 3.05$) at $1/n^2=0$ is the critical point obtained from the 
continuous time cluster algorithm. This value is compatible with the 
value $\Gamma_c\simeq 3.03$ obtained by extrapolation.  
}
\end{figure}

\begin{figure}
\epsfxsize=16cm\epsfysize=15cm\epsfbox{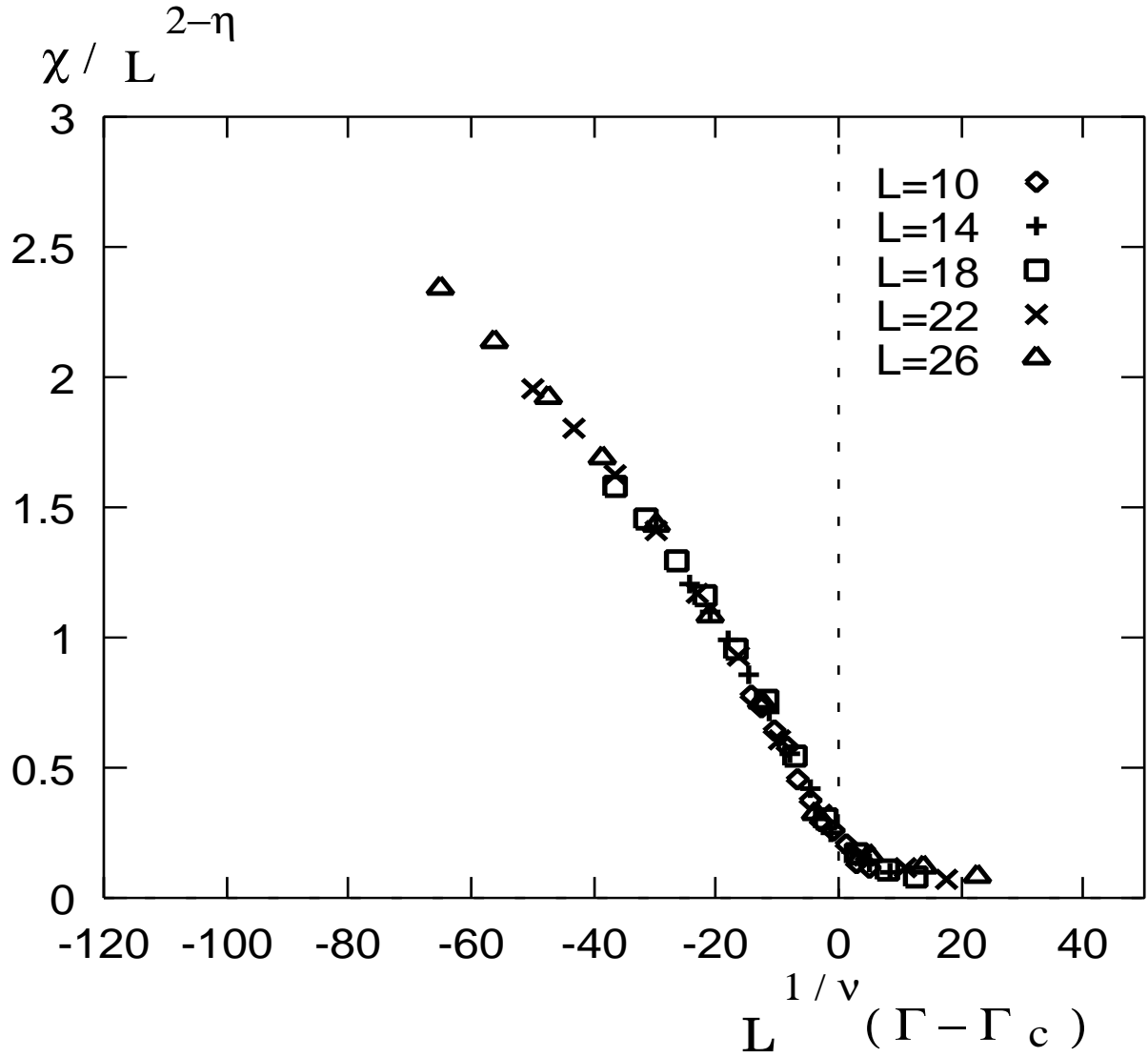}
\caption{
\label{fig4}
Scaled uniform susceptibility. In this case, $\Gamma_c=3.07$, $\nu=0.63$, and 
$\eta=0.04$. These estimates agree well with the series expansion results.}
\end{figure}

\begin{figure}
\epsfxsize=17cm\epsfysize=17cm\epsfbox{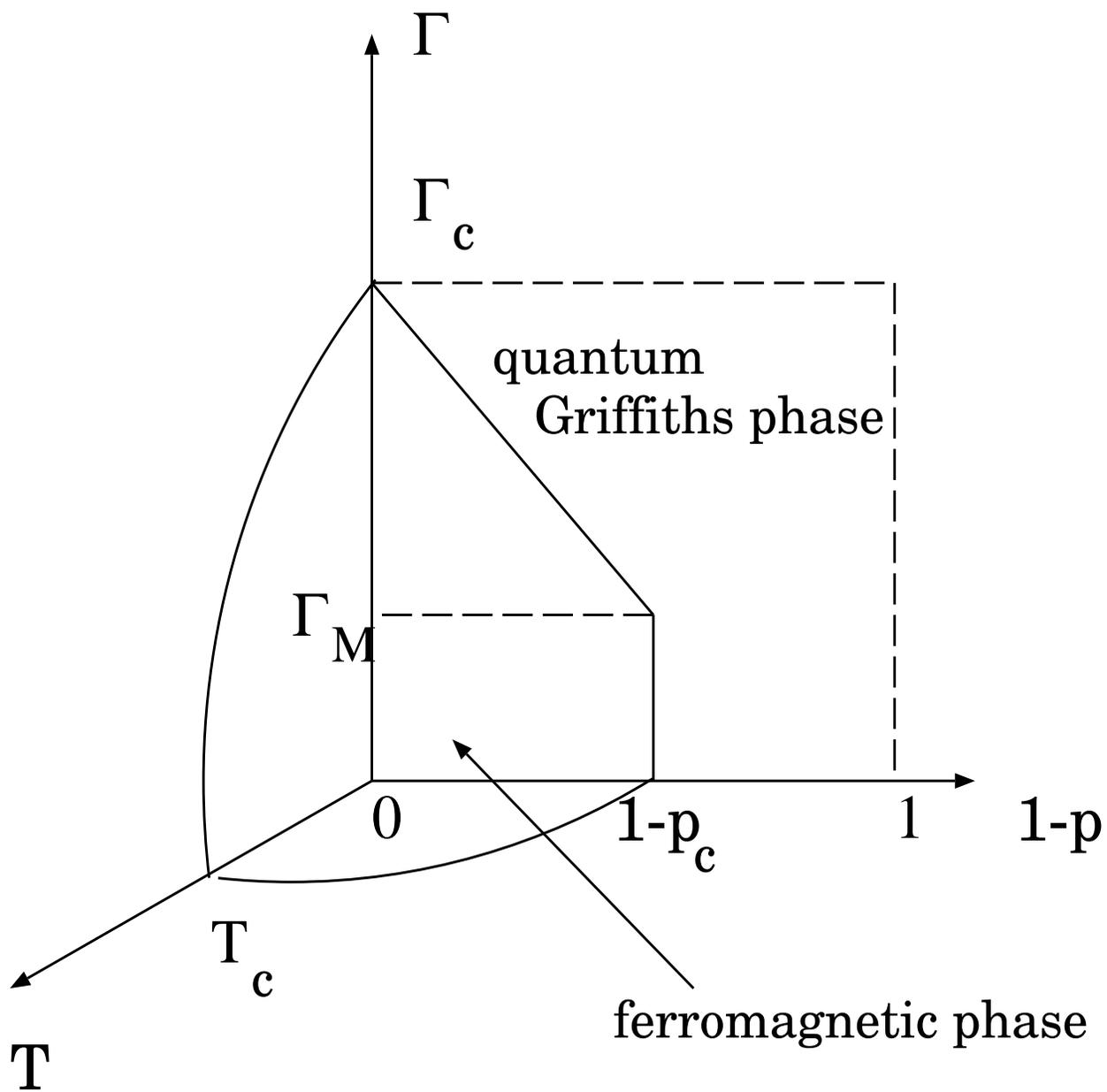}
\caption{
\label{fig5}
Schematic phase diagram of the site diluted system.
}
\end{figure}

\begin{figure}
\epsfxsize=16cm\epsfysize=16cm\epsfbox{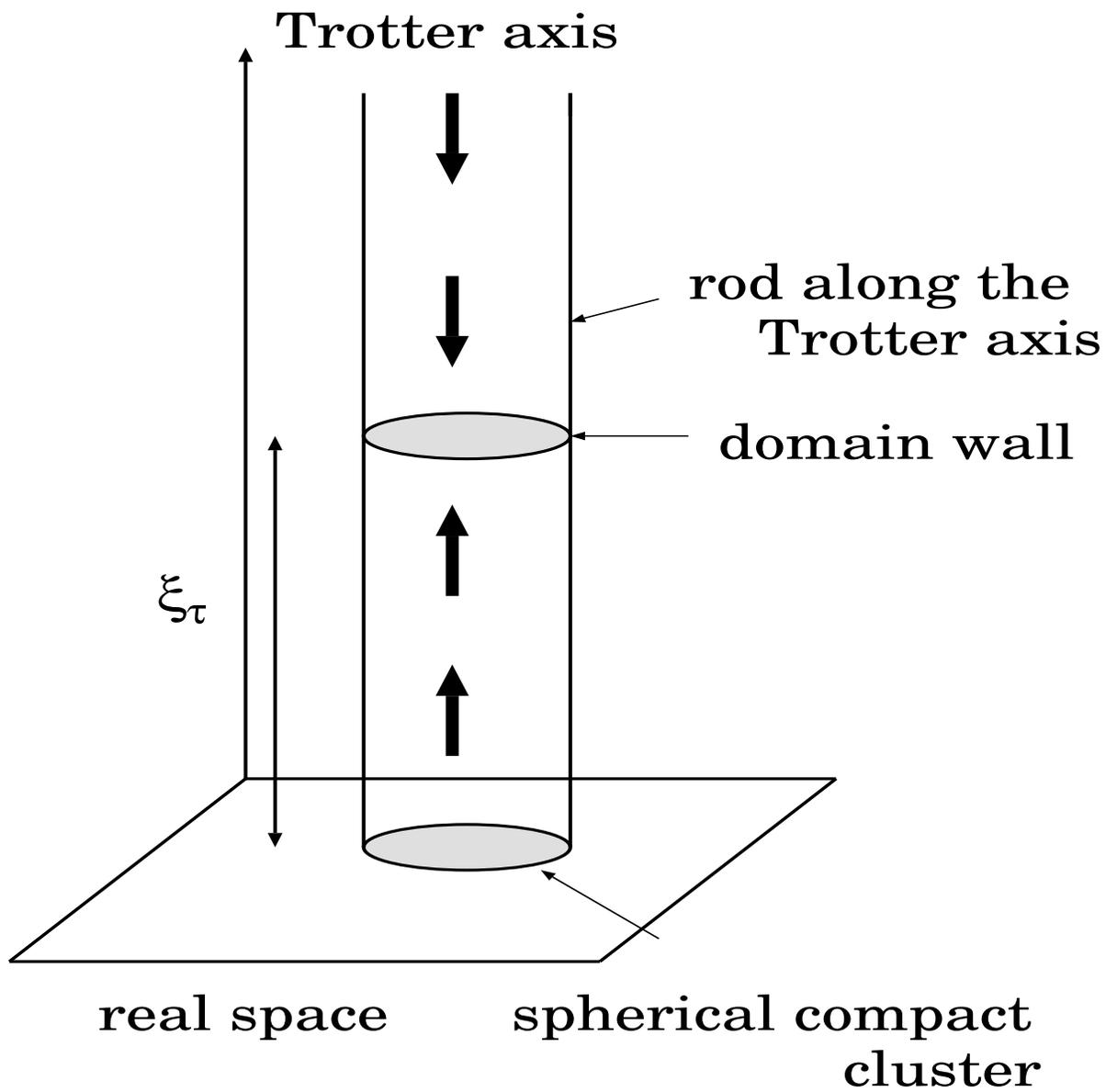}
\caption{
\label{fig6}
Schematic picture of a local cluster which forms a rod along the 
the Trotter axis.}
\end{figure}

\begin{figure}
\epsfxsize=16cm\epsfysize=18cm\epsfbox{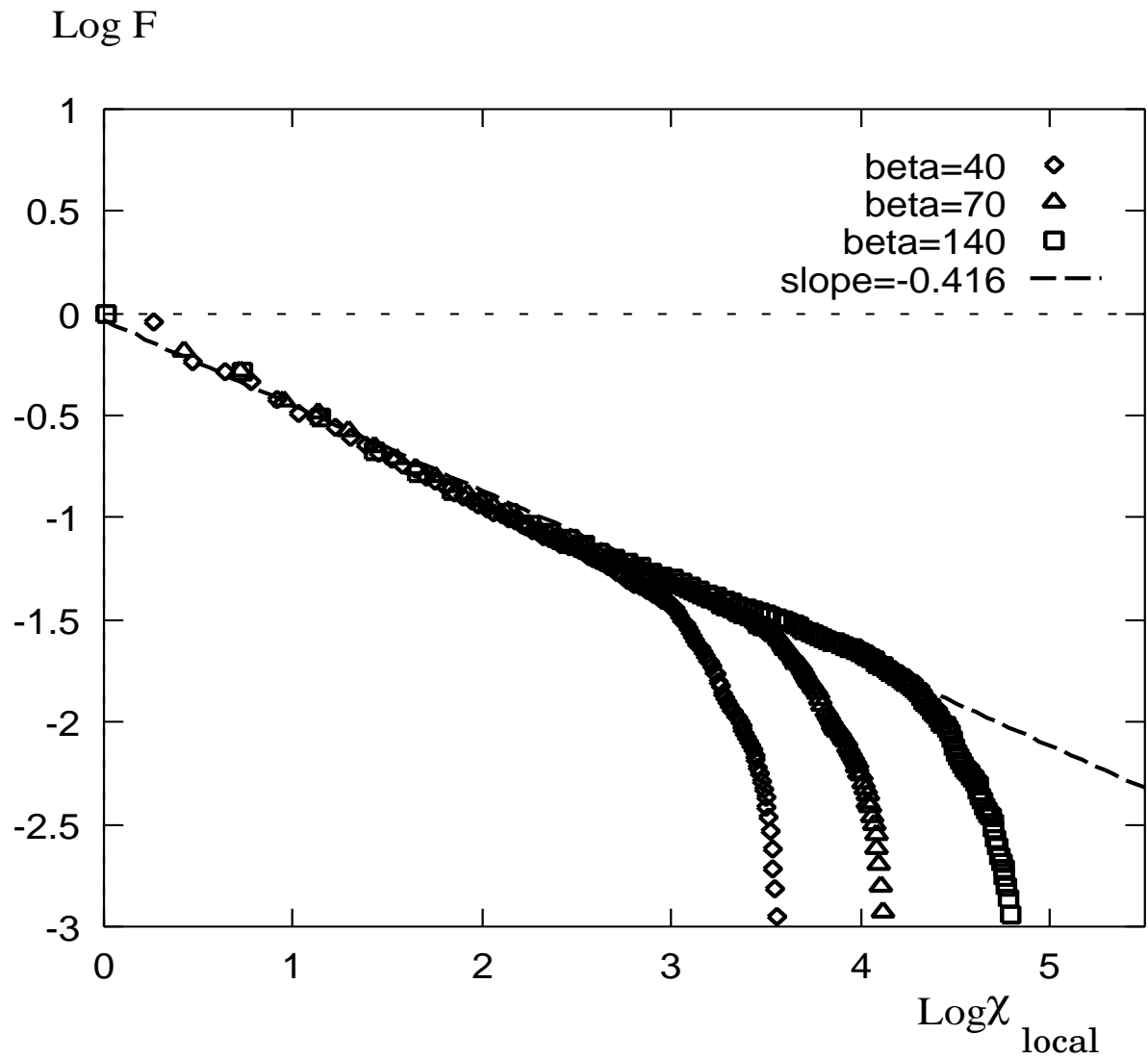}
\caption{
\label{fig7}
Log-Log plot of the integrated distribution of the single site susceptibility 
$\chi_{local}$ at $\Gamma=0.7$, $p=0.30$. In this case, we evaluate the 
distribution using $100$ different $20 \times 20$ configurations. The dotted 
line is a fit to the straight line region of the data and has the slope 
$-0.416$.}
\end{figure}

\begin{figure}
\epsfxsize=16cm\epsfysize=16cm\epsfbox{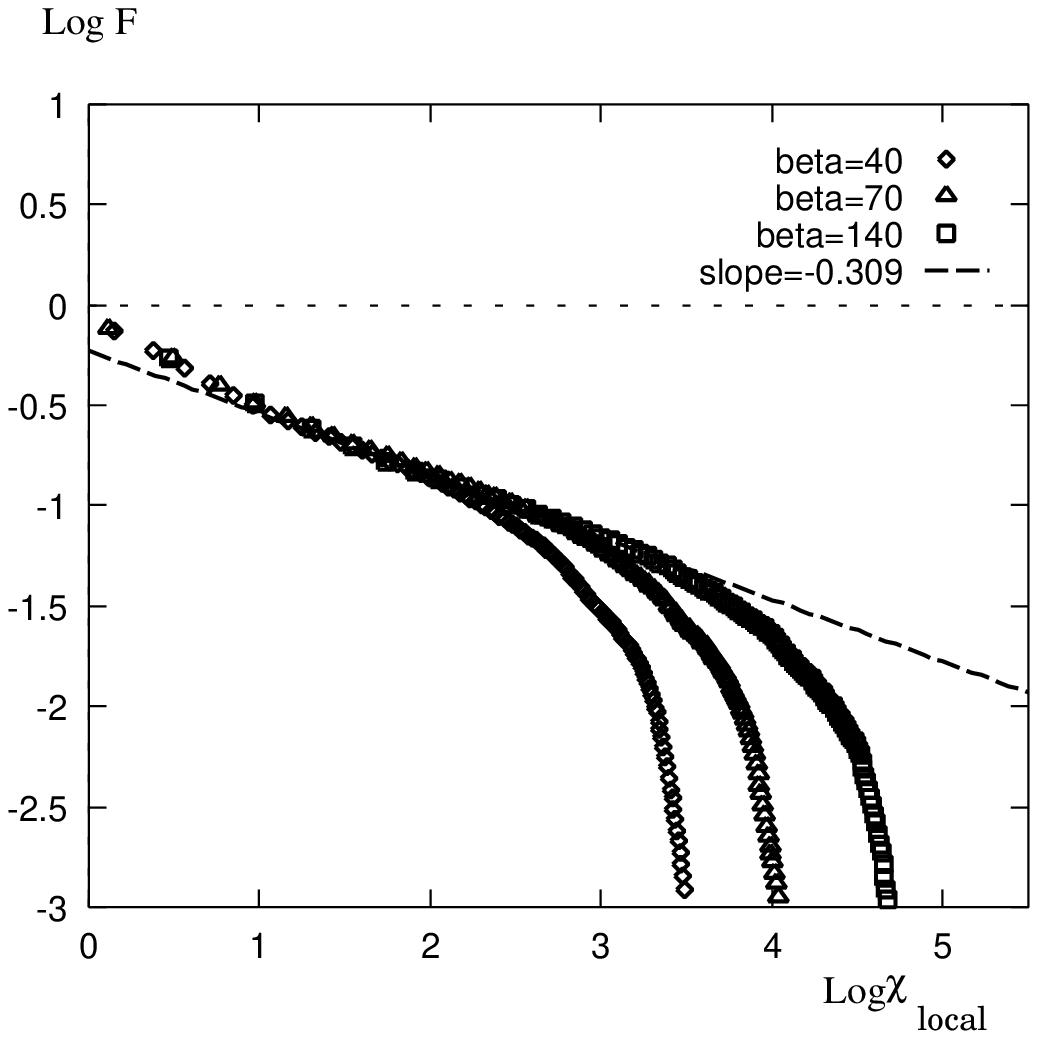}
\caption{
\label{fig8}
Similar to Fig. 6 but for $\Gamma=1.0$, $p=0.40$. In this case, we evaluate 
the distribution using $100$ different $20 \times 20$ configurations. 
The slope of the straight line is $-0.309$.}
\end{figure}

\begin{figure}
\epsfxsize=16cm\epsfysize=16cm\epsfbox{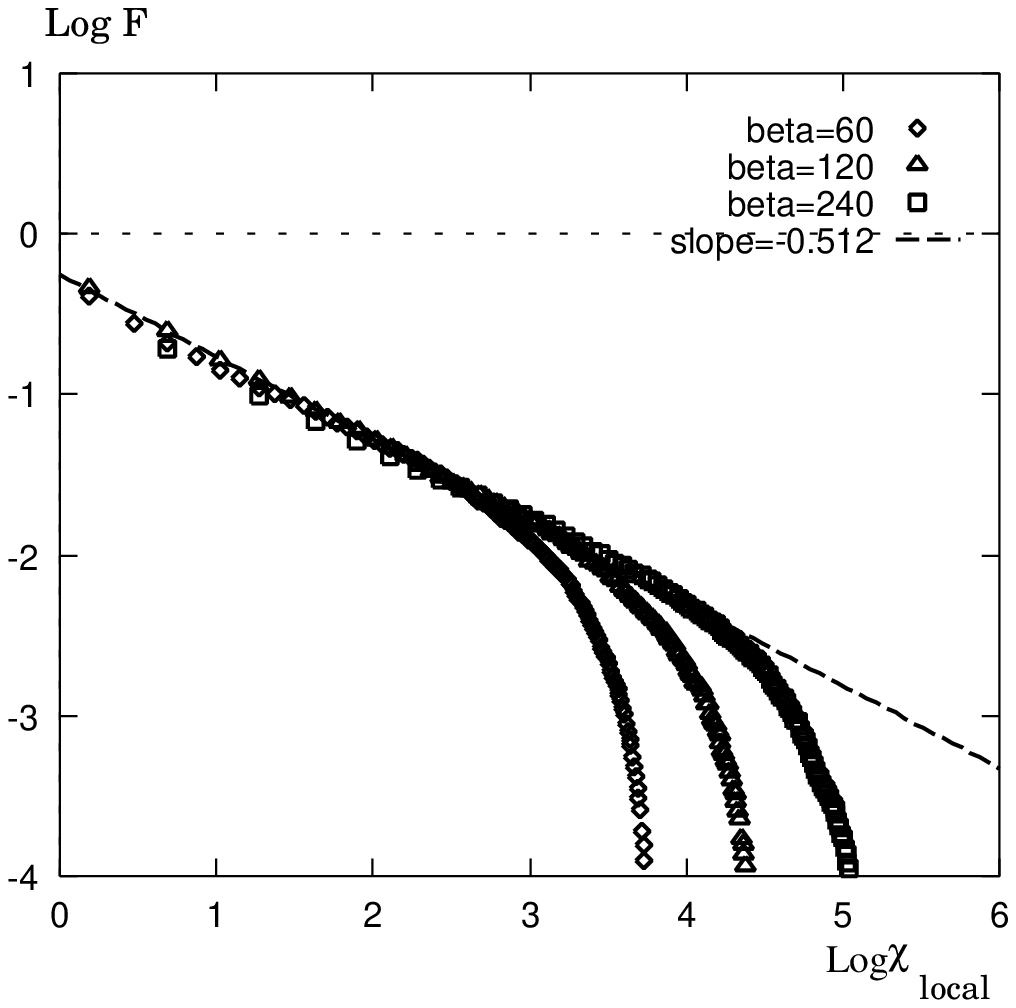}
\caption{
\label{fig9}
Similar to Fig. 6 but for $\Gamma=1.5$, $p=0.50$. In this case, we evaluate 
the distribution using $100$ different $20 \times 20$ configurations. 
The slope of the straight line is $-0.512$.}
\end{figure}

\begin{figure}
\epsfxsize=16cm\epsfysize=16cm\epsfbox{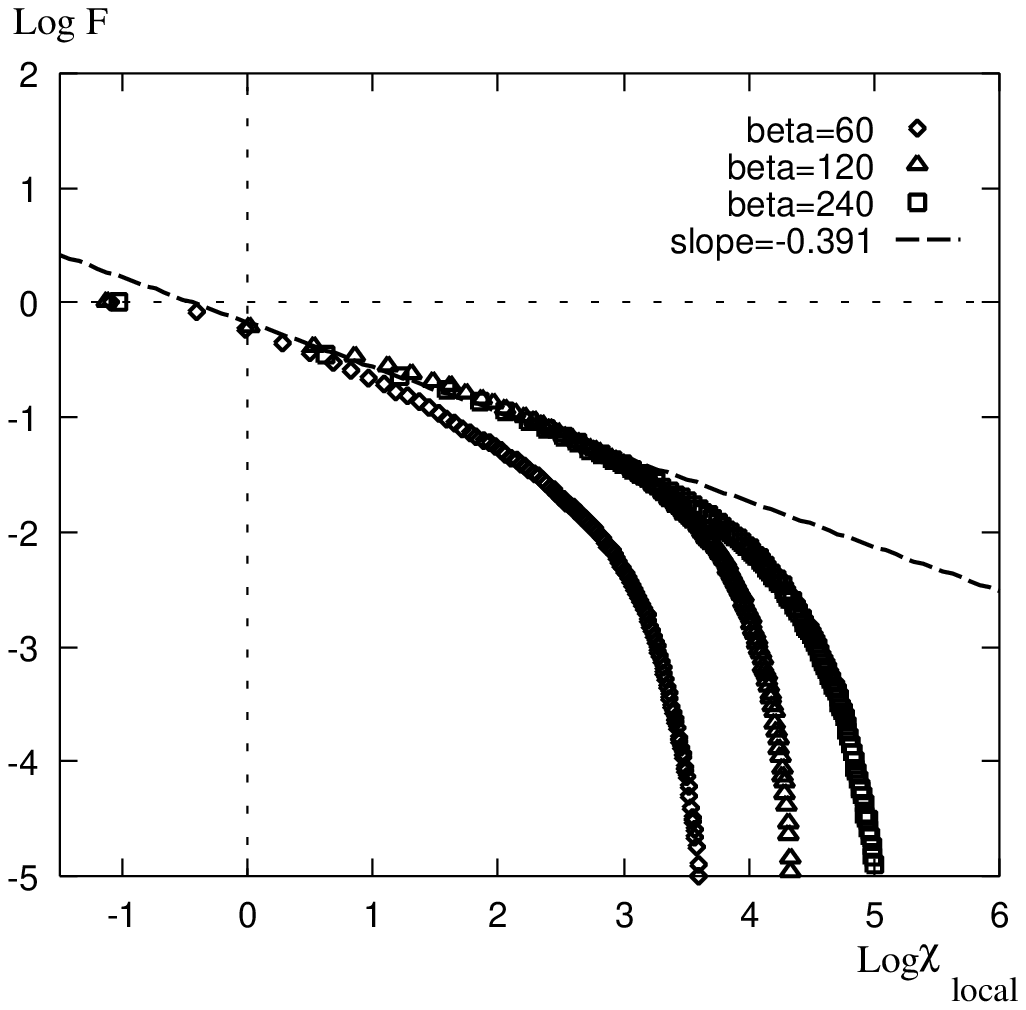}
\caption{
\label{fig10}
Similar to Fig. 6 but for $\Gamma=2.0$, $p=0.66$. In this case, we evaluate 
the distribution using $100$ different $20 \times 20$ configurations. 
The slope of the straight line is -0.391.}
\end{figure}

\begin{figure}
\epsfxsize=16cm\epsfysize=16cm\epsfbox{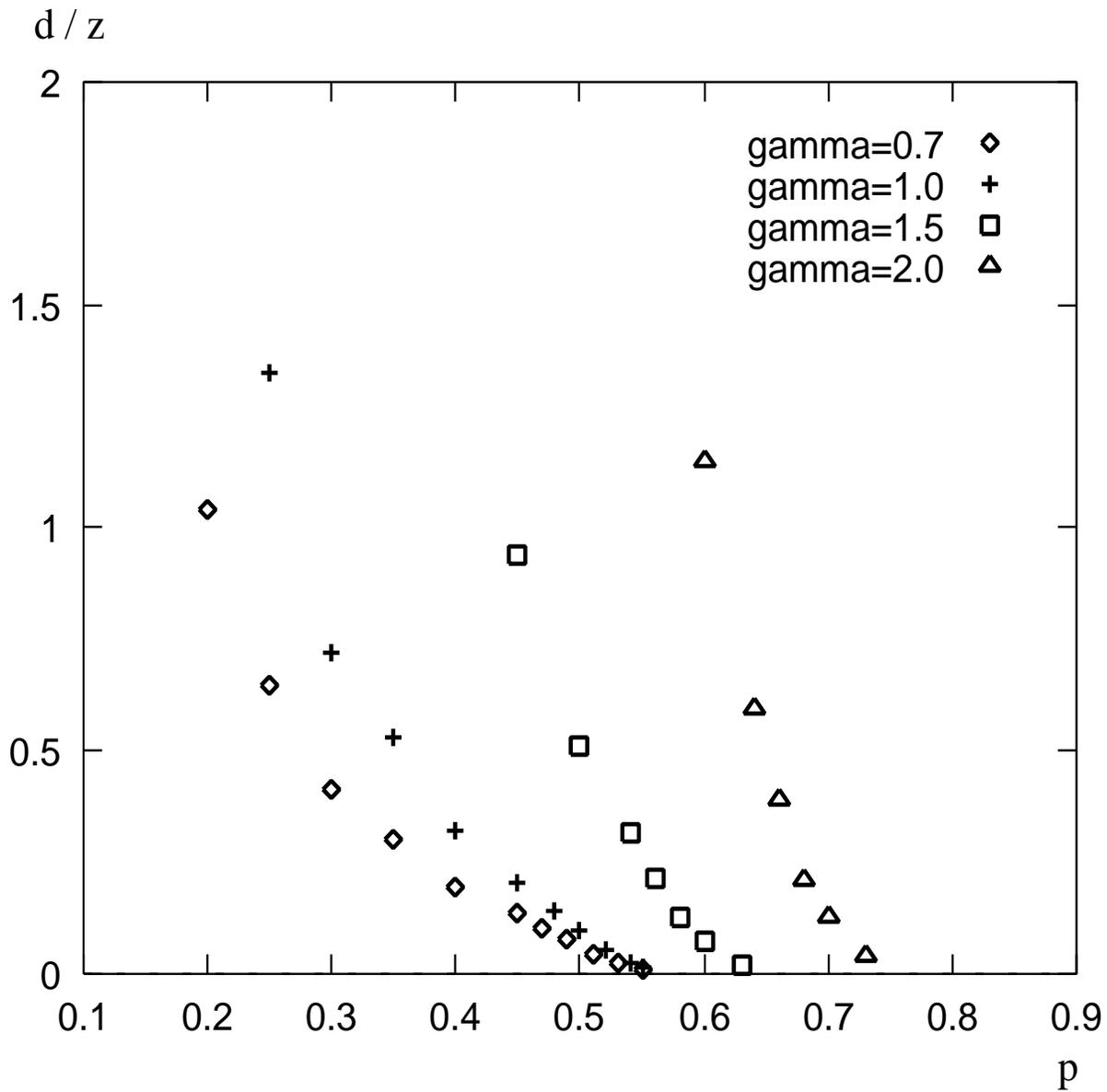}
\caption{
\label{fig11}
Plot of $d/z$ obtained by fitting the integrated distribution 
$F\left(\chi_{local}\right)$ for various $\Gamma$. For $\Gamma=0.7$ and 
$1.0$, $d/z$ vanish at $p\simeq p_c$, but, for $\Gamma=1.5$ and $2.0$, these 
vanish at $p > p_c$. When $z > d$, the average susceptibility will diverge.
}
\end{figure}

\begin{figure}
\epsfxsize=16cm\epsfysize=16cm\epsfbox{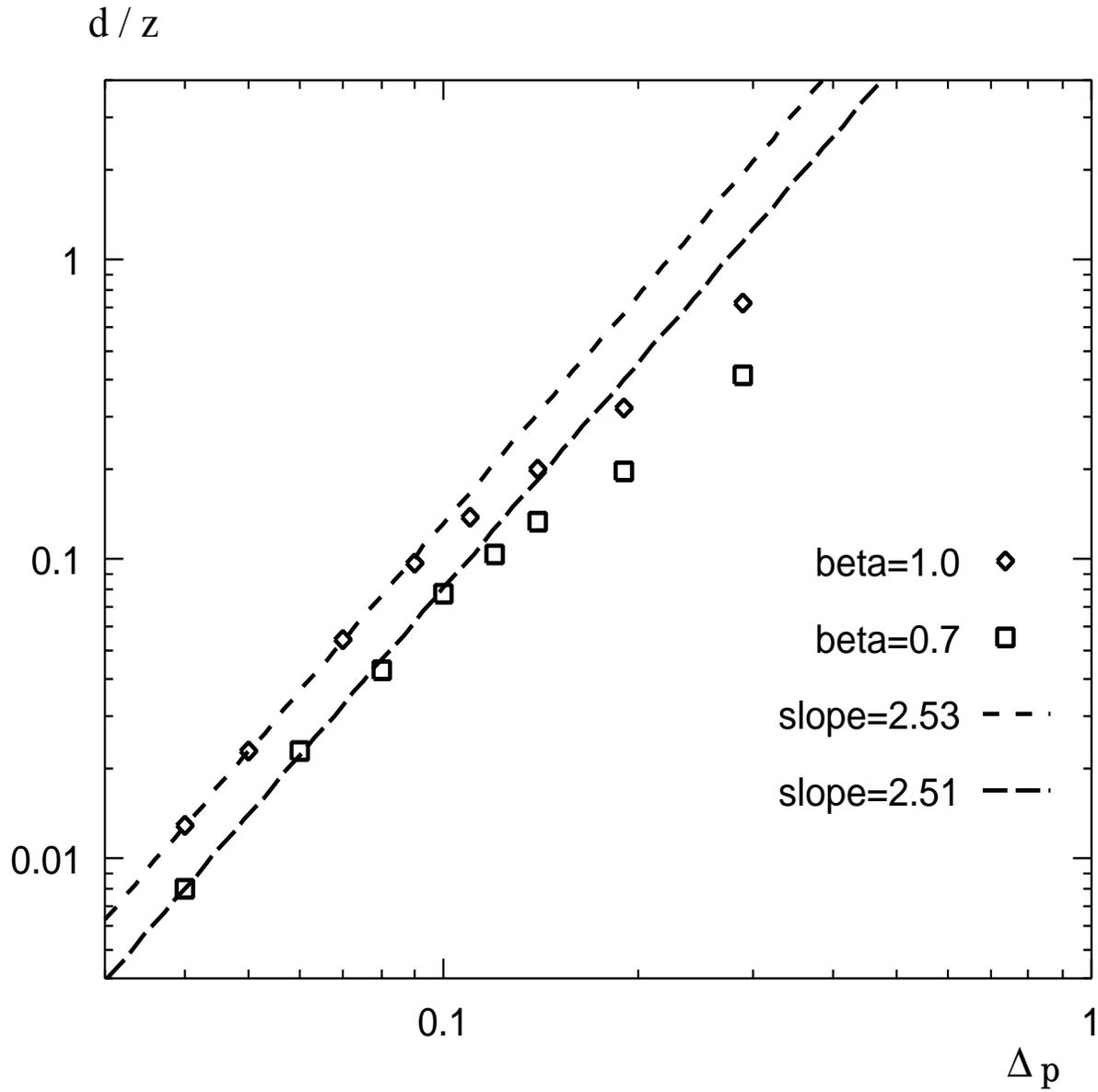}
\caption{
\label{fig12}
Extraction of the exponent from the data for $d/z$ for $\Gamma = 0.7$ and 
$1.0$, in which, $\Delta p$ implies $\vert p-p_c\vert$ ($p_c$ is $0.59$, 
the known percolation threshold). The value of $D\nu_p$ obtained from the 
percolation theory is known as $D\nu_p\simeq 2.57$, and the criticality of 
$d/z$ seems to be indeed compatible with the relation (21).
}
\end{figure}
\end{document}